\documentclass[twocolumn,prX]{revtex4}
\usepackage{graphicx}
\begin{document}
\title{Conformal Field Theory on the Fermi Surface}
\author{Brian Swingle}
\email{bswingle@mit.edu}
\affiliation{Department of Physics, Massachusetts Institute of Technology, Cambridge, MA 02139}
\begin{abstract}
The Fermi surface may be usefully viewed as a collection of $1+1$ dimensional chiral conformal field theories.  This approach permits straightforward calculation of many anomalous ground state properties of the Fermi gas including entanglement entropy and number fluctuations.  The $1+1$ dimensional picture also generalizes to finite temperature and the presence of interactions.  Finally, I argue that the low energy entanglement structure of Fermi liquid theory is universal, depending only on the geometry of the interacting Fermi surface.
\end{abstract}
\maketitle
\section{Introduction}
Fermi liquid theory forms the core of our theory of metals.  Many materials are well described by Fermi liquid theory over at least some portion of their phase diagram.  The experimental prominence and simplicity of Fermi liquids give them permanent appeal, so the community has expended much effort understanding their underlying structure.  The most physical way to understand the universality of Fermi liquids is in terms of renormalization group approaches that involve scaling towards the Fermi surface \cite{fermion_rg1,fermion_rg2,fermion_rg3}.  These approaches have given us a physical picture of Fermi liquids as mostly attractive fixed points.  To this mostly attractive fixed point we must add an infinite set of marginal deformations labeled by Landau parameters and the marginally relevant BCS instability.  The apparent stability of Fermi liquids has a dark side, however, as it has proved challenging to find simple systems exhibiting non-Fermi liquid behavior.  Indeed, Fermi liquids are continually surprising us with their versatility.

The latest surprise comes from the study of many body entanglement entropy.  Entanglement entropy is an attempt to characterize the real space entanglement properties of quantum ground states.  It is defined as the von Neumann entropy of the reduced density matrix of a spatial sub-system of the full system.  Most systems in $d_s > 1$ spatial dimensions satisfy a boundary law for the entanglement entropy of a spatial region \cite{arealaw1}, but free fermions violate this boundary law with a logarithmic correction \cite{fermion1,fermion2,fermion3,fermion4,fermion5,widom_proof}.  For a region of linear size $L$, the entanglement entropy of most known critical and non-critical systems is non-universal and scales as $S_L \sim L^{d_s - 1}$ \cite{arealaw1}.  However, the entanglement entropy for free fermions scales as $S_L \sim L^{d_s - 1} \ln{L}$ with the Fermi momentum $k_F$ making up the extra units where needed.  Furthermore, there is a precise conjecture for the form of this term known as the Widom formula \cite{fermion2}.  However, both the coefficient of the Widom formula and the extension to interacting fermions remain open questions.

The only other systems known to violate the boundary law are conformal field theories (or more generally scale invariant theories) in $d_s = 1$ spatial dimensions \cite{eeqft}.  I have suggested that these two violations of the boundary law are related because the Fermi surface in any dimension may be regarded as a collection of $1+1$ dimensional chiral conformal field theories \cite{bgs_ferm1}.  However, it should be understood that this correspondence, at least in its simplest form, is only expected to hold in the low energy limit.  I also wish to emphasise that this $1+1$ dimensional construction is not an attempt to bosonize the Fermi surface in higher dimensions.  Note that a more complicated patching procedure has recently been employed to discuss a Fermi surface coupled to a gapless boson, but I will not address these issues here \cite{fs_boson1,fs_boson2}.

In this paper I give a more complete formulation of a Fermi surface as a collection of $1+1$ dimensional chiral conformal field theories.  I will first describe the basic setup for free fermions and sketch the arguments leading to the anomalous entanglement entropy of the free Fermi gas.  As examples of the formalism, I compute number fluctuations and heat capacity of a Fermi gas from the $1+1$ dimensional point of view.  Finally, I include interactions and argue that interacting Fermi liquids violate the boundary law for entanglement entropy in a universal way.  I conclude with some comments about the relevance of these results to other systems and about future work.

\section{Chiral Fermions on the Fermi Surface}
We wish to understand in what sense a Fermi surface can be described as a collection of $1+1$ dimensional chiral conformal field theories.  I consider fermions on a lattice, although much of what I say is independent of the details of the ultraviolet regulator.  Given a generic band structure and filling fraction, a finite density of fermions will form a metallic state with a Fermi surface.  For simplicity, let us assume that this Fermi surface is nearly spherical.  The low energy degrees of freedom are particle-hole fluctuations near the Fermi surface in momentum space.  Note that quasiparticles exist only above the Fermi surface while quasiholes exist only below.  Both particles and holes move with the same group velocity set by the local Fermi velocity.

Each such ``patch" on the Fermi surface is equivalent to a single gapless chiral fermion in $1+1$ dimensions, the dimensions being the radial direction and time.  The patch is chiral because all the excitations have a common velocity.  For illustrative purposes, let us specialize to the case of $d_s = 2$ so that the Fermi surface is one dimensional.  We can formalize these statements by saying that the low energy effective action of the free Fermi gas is
\begin{equation}
\mathcal{S}_{\psi} = \int d\theta \int \,dk\, dt \, \psi^+_\theta(k,t) [i\partial_t - v_F k ] \psi_\theta(k,t),
\end{equation}
where $\theta$ labels the patch on the Fermi surface as in Fig. 1. Each operator $\psi_{\theta}(k,t)$ should be regarded as a free fermion in one spatial dimension, the local radial direction, as specified by $\theta$. The basic approach will be to compute physical properties of the Fermi gas by appropriate sums over the $1+1$ dimensional degrees of freedom labeled by $\theta$.

\begin{figure}
\hspace*{-180pt}
\includegraphics[width=.75\textwidth]{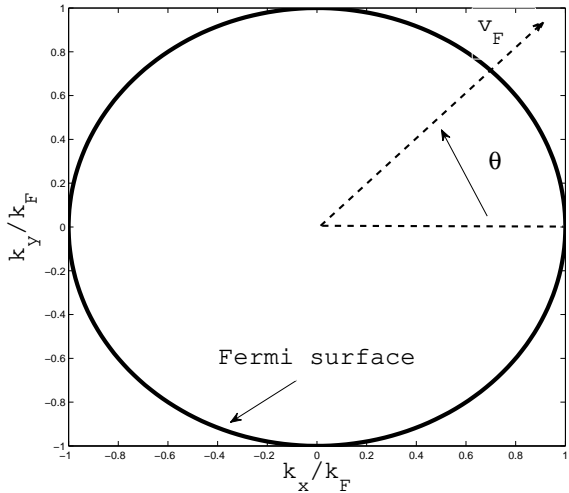}
\label{fig1}
\hspace*{-180pt}
\caption{A sketch of a circular Fermi surface in $d_s = 2$ dimensions.  The angle $\theta$ parameterizes the Fermi surface and labels the local patches.  The Fermi velocity for a particular $\theta$ is shown.}
\end{figure}

Starting from this free low energy effective action, we can trace the effects of interactions using a renormalization group flow, but I will continue to focus on the case of free fermions.  Each patch, labeled by $\theta$, is equivalent to the chiral half of a one dimensional free relativistic fermion.  The non-chiral relativistic fermion has a total central charge $c = 2$ as a conformal field theory.  This central charge is a sum of left and right moving pieces $c = c_L + c_R$ where for the free relativistic fermion we have $c_L = c_R = 1$.  This assignment should not be confused with the Majorana fermion which has central charges $c_L = c_R = 1/2$.  The basic fact we will use about such a one dimensional conformal field theory is that the entanglement entropy on an interval of length $L$ is given by $S = \frac{c_L + c_R}{6} \ln{(L/\epsilon)}$ where $\epsilon$ is an ultraviolet cutoff.  We will also use the generalization of this formula to finite temperature $T = 1/\beta$ given by
\begin{equation}
S = \frac{c_L + c_R}{6} \ln{\left(\frac{\beta v}{\pi \epsilon} \sinh{\left(\frac{\pi L}{\beta v}\right)}\right)}
\end{equation}
where $v$ is the characteristic velocity in the conformal field theory \cite{eeqft}.  Later $v$ will be identified with the local renormalized Fermi velocity $v_F$.

The formulation sketched above gives an intuitive understanding of the anomalous entanglement properties of the Fermi surface.  If we ask about the entanglement entropy in a region $A$ of linear size $L$, then we naturally coarse grain the Fermi surface into ``coarse patches" of typical size $1/L^{d_s - 1}$ for a total of $ (k_F L)^{d_s - 1}$ ``coarse patches".  Each patch contributes roughly $\ln{L}$ as is appropriate for a one dimensional conformal field theory, and the total entanglement entropy scales as $L^{d_s -1} \ln{L} $ as observed \cite{fermion1,fermion2,fermion3,fermion4,fermion5}.  The Widom formula for the entanglement entropy is obtained from a more precise counting of patches \cite{bgs_ferm1}.  The fundamental interpretation of this counting procedure and the Widom formula is in terms of an effective central charge coming from the amalgamation of many $1+1$ dimensional degrees of freedom \cite{bgs_ferm1}.  We will now use similar counting arguments to compute a number of observables for free fermions.

\section{Number Fluctuations at Zero Temperature}
In addition to anomalous entanglement entropy, the free Fermi gas has anomalously large fluctuations of some conserved quantities.  Similar to the story of entanglement entropy, most systems in $d_s$ spatial dimensions satisfy a boundary law for ground state fluctuations of various physical quantities. As an example, let us consider the number operator $N_A$ corresponding to the number of fermions in region $A$.  Typically, one would expect to find $\langle (N_A - \langle N_A \rangle )^2\rangle \sim L^{d_s -1}$ where $L$ is the linear size of region $A$.  This boundary law for fluctuations in the ground state appears to be relatively universal, however, it is violated in the case of free fermions \cite{fermion2,widom_proof}.  Just like their entanglement properties, free fermions have anomalously large number fluctuations scaling like $L^{d_s - 1} \ln{L} $.  It is natural to ask if we can account for these fluctuations by viewing the Fermi surface as a collection of chiral one dimensional conformal field theories.

The leading logarithmically corrected boundary law behavior can again be traced to the presence of numerous gapless modes at the Fermi surface.  Consider first the problem of number fluctuations in a one dimensional gas of free non-relativistic fermions at finite density.  We wish to calculate the fluctuations $\Delta N^2_A = \langle (N_A - \langle N_A \rangle )^2\rangle$ in the fermionic ground state with $A$ an interval of length $L$.  By the writing the operator $N_A$ as a restricted integral over the fermion density operator the calculation can be reduced to an integral using Wick's theorem.  To leading order in $L$ the fluctuations scale as $\Delta N_A^2 \sim \frac{1}{\pi^2}\ln{(L/\epsilon)}$.  Note that this result generalizes to Luttinger liquids in one dimension \cite{fs_fluc_1d}.  There are two Fermi points, or patches, and each Fermi point contributes $\frac{1}{2}\frac{1}{\pi^2}\ln{L}$ to the answer.  Returning to $d_s > 1$ dimensions, the number fluctuations can again be written in integral form using Wick's theorem.  However, the analysis of the integral is considerably more complex \cite{widom_proof,num_fluc}.  Instead, we can obtain the exact expression for the asymptotic behavior of the number fluctuations indirectly using the one dimensional picture.

To perform the mode counting, let us return for a moment to the entanglement entropy.  We choose a spatial region $A$ of linear size $L$ and ask about the entanglement entropy of this region.  The Widom formula takes the form of an integral over the boundary of $A$ and the Fermi surface
\begin{equation}
S = \frac{1}{(2\pi)^{d_s -1}}\frac{\ln{L}}{12} \int_k \int_x \, dA_k dA_x |n_x \cdot n_k |
\end{equation}
where $n_x$ and $n_k$ stand for unit normals to the boundary of $A$ and the Fermi surface respectively.  The detailed choice of linear size $L$ in the logarithm is immaterial in this formula since it only corrects the ultraviolet sensitive boundary law piece of the entanglement entropy.  Once more, this formula should be interpreted as counting the effective number of chiral one dimensional modes contributing $\ln{L}$ to the entropy.  Each patch has $c_L = 1$ and $c_R = 0$ where left and right movers are defined by the local radial direction.  Thus each patch should contribute $\frac{c_L + c_R}{6} \ln{L} = \frac{1}{6} \ln{L}$ to the entropy.  The Widom formula is essentially this entropy per patch times the number of such patches
\begin{equation}
N_{modes} = \frac{1}{(2\pi)^{d_s -1}}\frac{1}{2} \int_k \int_x \, dA_k dA_x |n_x \cdot n_k |
\end{equation}
where the $|n_x \cdot n_k|$ factor arises from computing the projected area when doing the mode counting \cite{bgs_ferm1}.

Returning to the number fluctuations, each patch contributes $\frac{1}{2}\frac{1}{\pi^2}\ln{L}$ following the one dimensional result, and the asymptotic form of the number fluctuations for a region $A$ in higher dimensions is thus
\begin{equation}
\Delta N_A^2  = \frac{1}{(2\pi)^{d_s -1}}\frac{\ln{L}}{4 \pi^2} \int_k \int_x \, dA_k dA_x |n_x \cdot n_k | .
\end{equation}
Precisely this formula has been obtained previously by a lengthier and rigorous analysis (note that some confusion can arise in the comparison because of different bases for the logarithms) \cite{fermion2,widom_proof}.

\section{Generalization to Finite Temperature and Interactions}
So far we have worked at zero temperature and without interactions, but these restrictions are not essential.  For example, the heuristic picture of the Fermi surface as a collection of one dimesional chiral patches gives roughly the correct finite temperature entropy for a region $A$ of linear size $L$.  Each patch contributes roughly $L T$ to the entropy and the number of patches is proportional $(k_F L)^{d_s - 1}$ giving roughly the correct entropy for free fermions.  There is a reason to be skeptical however.  The details of the patch counting depend on the boundary of region $A$, but any such dependence should disappear in thermodynamic quantities.  Fortunately, the physically correct choice of the linear size $L$ naturally cancels this detailed boundary dependence.

\begin{figure}
\includegraphics[width=.5\textwidth]{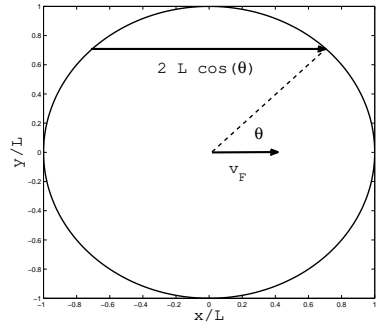}
\label{fig2}
\caption{A sketch of a circular real space region $A$ of radius $L$.  The Fermi velocity of a particular patch on the Fermi surface is shown superimposed on the circle.  The effective length $2 L \cos{\theta}$ is shown as a function of angle $\theta$ relative to the direction defined by $v_F$.}
\end{figure}

Returning to the case $d_s = 2$, let's assume that the Fermi surface is a circle of radius $k_F$.  The real space region $A$ is taken to be a circle of radius $R$.  Consider a specific patch on the Fermi surface and choose coordinates so that this patch has velocity equal to $v_F \hat{x}$ as in Fig. 2.  To compute the integral over the boundary of region $A$ we must specify what the effective linear size $L$ is.  We parameterize the boundary of $A$ by an angle $\theta$ defined relative to the x-axis.  For a mode with velocity $v_F \hat{x}$ and a point on the boundary of $A$ labeled by $\theta$, the physically correct linear size is $L_{\text{eff}} = 2L | \cos{\theta} |$, that is the chordal distance across the circle parallel to the x-axis as in Fig. 2.  The high temperature limit of Eq. 2 with $L = L_{\text{eff}}$ is
\begin{equation}
S_{1+1} = \frac{c_L + c_R}{6} \frac{\pi L_{\text{eff}}}{\beta v_F}.
\end{equation}
We use this expression for the entropy contribution of each patch to compute the entropy of the entire Fermi surface \begin{equation}
S = \frac{1}{2\pi}\frac{1}{2} \int_k \int_x \, dA_k dA_x |n_x \cdot n_k | \frac{1}{6} \frac{\pi L_{\text{eff}}}{\beta v_F} .
\end{equation}
The integral over the real space boundary can be carried out as described above.  What remains is an integral of $1/v_F$ over the Fermi surface, a familiar result giving the density of states.  The final result is
\begin{equation}
S = \frac{\pi}{6} m T (\pi L^2)
\end{equation}
where $m$ is the fermion mass and $\pi L^2$ is the area of region $A$.  This is nothing but the usual thermal entropy of a gas of free fermions in a box of area $\pi L^2$ to leading order in $T/T_F$.  A similar calculation gives the correct finite temperature number fluctuations to leading order in $T/T_F$.  The fact that the Widom formula, and the patch counting it embodies, can be used to compute exact low temperature thermal properties is strong evidence in its favor.

It is also possible to include interactions.  In the renormalization group treatment of Fermi liquids, most interactions are irrelevant.  The exceptions are forward scatting interactions, but these interactions do not drastically modify the $1+1$ dimensional picture.  The low energy effective action Eq. 1 has an emergent $U(1)^\infty$ symmetry corresponding to number conservation on each patch, and this symmetry survives in the low energy limit of Fermi liquids because only forward scattering remains.  Sticking with the simplest possible situation, let us assume that approximate rotational symmetry is preserved as we turn on interactions.  With the assumption of rotational symmetry, Luttinger's theorem that the area (in $d_s = 2$ dimensions) enclosed by the Fermi surface remains constant implies that $k_F$ is not renormalized by interactions.  Indeed, the content of Fermi liquid theory is that the effects of interactions may be subsumed entirely in terms of a renormalized Fermi velocity and a set of Landau parameters.

To see that interactions can be automatically included, let us compute the heat capacity of a Fermi liquid.  The standard result of Fermi liquid theory is that the heat capacity depends on interactions only through the renormalized mass (assuming $k_F$ remains at its free value).  Remarkably, we can describe the thermodynamics of an interacting Fermi liquid, say in $d_s = 2$, using Eq. 7 so long as we replace the bare Fermi velocity by the physical renormalized Fermi velocity.  In particular, the mode counting remains unchanged despite the presence of interactions.  Given that we can reproduce thermodynamics exactly using the patch construction, we have a strong hint that the interacting Fermi liquid does indeed have an entanglement entropy described by the Widom formula.  Dropping the assumption of spherical symmetry, the formalism predicts that the entanglement entropy is still highly universal, it is totally insenstive to the Landau parameters, for example, and depends only on the shape of the interacting Fermi surface.

\section{Discussion and Conclusions}
I have elaborated on a view of the Fermi surface as a collection of $1+1$ dimensional chiral conformal field theories. I focused mostly on circular real space regions and Fermi surfaces in $d_s = 2$ spatial dimensions, but the generalization to other geometries and higher dimensions is straightforward.  So long as the region $A$ is convex, the proper effective length is physically well defined.  To define this length, draw a line from a fixed point on the spatial boundary such that the line is parallel to the Fermi velocity of a fixed point on the Fermi surface.  For convex $A$ this line will intersect the boundary of $A$ in one other place.  The length of the line segment between these two intersections is $L_{\text{eff}}$ for the chosen points.  Using this definition, the thermal entropy calculation can be shown to produce the volume of the region $A$ irrespective of the particular shape chosen.  The ultimate message is that the anomalous entanglement entropy depends only on the geometry of the interacting Fermi surface in any dimension.  This result is part of a growing body of work that strongly links entanglement and geometry in many body physics.

The present formulation was originally motivated by renormalization group treatments of the Fermi liquid and by attempts to understand intuitively the origin of the anomalous entanglement properties of fermions.  However, it seems to giving us much more.  It can handle both finite temperature and some kinds of interactions, and it provides strong evidence both for the still unproven Widom formula for free fermions and for its extension to interacting fermions.  On the practical side, it provides a simple and unified formalism for many calculations that previously appeared quite complex in the free Fermi gas.

These results also support the prediction that other exotic phases of matter, including non-Fermi liquid metals \cite{crit_fs}, spin liquids with a spinon Fermi surface, holographic non-Fermi liquids \cite{nfl1}, and Bose metals \cite{dbl} violate the boundary law for entanglement entropy.  To truly extend the patch argument given here to these systems, one must account for non-trivial coupling between patches such as that provided by a gapless boson, but I leave this to future work.  It is also likely that these exotic phases have anomalously large fluctuations analogous to the fermion number fluctuations in a Fermi liquid, although these statements are more model dependent.  One hope is that anomalous properties like entanglement entropy and fluctuations may provide useful numerical and experimental handles for identifying such exotic phases.  Finally, this work suggests a generalization of Fermi liquid theory to produce non-Fermi liquids by putting other chiral conformal field theories on the Fermi surface \cite{chiral_fs}.

\section{Acknowledgements}
I would like to thank M. Barkeshli and T. Grover for many discussions, N. Bray-Ali for discussions about universality, and V. Kumar and J. McGreevy for comments on the manuscript.  I would also like to thank Xiao-Gang Wen for support and encouragement during this work.

\bibliography{cft_ff}
\end{document}